\documentclass[
   ,draft            
   ,numberedheadings 
  ]
  {aipproc}

\layoutstyle{8x11single}

\begin{document}

\title{Status of Chiral Meson Physics}


\author{Johan Bijnens}{
  address={Department of Astronomy and Theoretical Physics, Lund University, S\"olvegatan 14A, SE 22362 Lund, Sweden}
}

\begin{abstract}
This talk includes a short introduction to Chiral Perturbation Theory
in the meson sector concentrating on a number of recent developments.
I discuss the latest fit of the low-energy constants.
Finite volume corrections are discussed for the case with
twisted boundary conditions for form-factors
and first results at two-loops for three flavours for masses.
The last part discusses the extension to other symmetry breaking patterns
relevant for technicolour and related theories as well as the
calculation of leading logarithms to high loop orders.
\end{abstract}

\renewcommand{\today}{} 
\begin{titlepage}
\begin{flushright}
\fontsize{12}{12}\selectfont
LU-TP 14-37\\[2mm]
November 2014
\end{flushright}
\vfill
\begin{center}
{\fontsize{20}{20}\bfseries Status of Chiral Meson Physics$^\dagger$}
\vfill
{\fontsize{14}{14}\bfseries Johan Bijnens}\\[0.8cm]
{\fontsize{12}{12}\normalfont Department of Astronomy and Theoretical Physics, Lund University,\\[2mm]
S\"olvegatan 14A, SE 223-62 Lund, Sweden}
\vfill
\begin{minipage}{0.8\textwidth}
\begin{center}
\bf Abstract
\end{center}
This talk includes a short introduction to Chiral Perturbation Theory
in the meson sector concentrating on a number of recent developments.
I discuss the latest fit of the low-energy constants.
Finite volume corrections are discussed for the case with
twisted boundary conditions for form-factors
and first results at two-loops for three flavours for masses.
The last part discusses the extension to other symmetry breaking patterns
relevant for technicolour and related theories as well as the
calculation of leading logarithms to high loop orders.
\end{minipage}
\end{center}
\vfill
\noindent $^\dagger$ Plenary talk at ``XIth Quark Confinement and the Hadron
 Spectrum, Saint-Petersburg, Russia, 8-12 September 2014.
\end{titlepage}

\maketitle

\section{Introduction}

This talk is an overview of mesonic Chiral Perturbation Theory (ChPT)
concentrating on a number of recent developments.

\section{Chiral Perturbation Theory}

Chiral Perturbation Theory can best be described by ``Exploring the
consequences of the chiral symmetry of QCD and its spontaneous breaking
using effective field theory techniques.''
It was introduced by Weinberg, Gasser and Leutwyler
\cite{Weinberg:1978kz,Gasser:1983yg,Gasser:1984gg}. A good discussion
of the underlying assumptions can be found in~\cite{Leutwyler:1993iq}.
References to lectures and other material can be found in~\cite{chptwebpage}.

A general effective field theory (EFT) needs three principles: 
the correct degrees of freedom, there has to be a power-counting
principle to ensure predictivity and one should remember the associated range
of validity. For ChPT the degrees of freedom are the Nambu-Goldstone bosons
from the spontaneous symmetry breaking of the chiral symmetry of massless
QCD, the power-counting for mesonic ChPT is dimensional counting in momenta
and meson masses, and the range of validity stops somewhere below the mass
of the first not included resonance, the rho.

The QCD Lagrangian
\begin{equation}
{\cal L}_{QCD} =  \sum_{q=u,d,s}
\left[i \bar q_L D\hskip-1.3ex/\, q_L +i \bar q_R D\hskip-1.3ex/\, q_R
- m_q\left(\bar q_R q_L + \bar q_L q_R \right)
\right]
\end{equation}
has an $SU(3)_L\times SU(3)_R$ global chiral symmetry when $m_q=0$. This
symmetry is spontaneously broken by the quark-antiquark
vacuum-expectation-value $\langle\overline q q \rangle=
\langle\overline q_L q_R \overline q_R q_L \rangle\ne 0$. The mechanism is
discussed in the talk by L.~Giusti. The remaining symmetry group is
$SU(3)_V$, we have thus 8 broken generators and get 8 Goldstone
bosons whose interaction vanishes at zero momentum. The latter allows for
a consistent power counting via dimensional counting \cite{Weinberg:1978kz}.

There are many extensions of ChPT in different directions. Some of them are:
\begin{itemize}
\item Which chiral symmetry:
  $SU(N_f)_L\times SU(N_f)_R$, for $N_f=2,3,\ldots$ and
  extensions to (partially) quenched
\item Or beyond QCD
\item Space-time symmetry:
  Continuum or broken on the lattice:
    Wilson, staggered, mixed action
\item Volume: Infinite, finite in space, finite T
\item Which interactions to include beyond the strong one
\item Which particles included as non Goldstone Bosons
\end{itemize}
My general belief is that if it involves soft pions (or soft $K,\eta$) some
version of ChPT exists.

The Lagrangians are written in terms of the special unitary matrix,
parametrizing $SU(3)_\times SU(3)_R/SU(3)_V\approx SU(3)$,
\begin{equation}
U=e^{i\sqrt{2}\Phi/F_0}\quad\mathrm{with}\quad
\Phi (x) = \, \left( \begin{array}{ccc}
\displaystyle\frac{ \pi^0}{ \sqrt 2} \, + \, \frac{ \eta_8}{ \sqrt 6}
 & \pi^+ & K^+ \\
\pi^- &\displaystyle - \frac{\pi^0}{\sqrt 2} \, + \, \frac{ \eta_8}
{\sqrt 6}    & K^0 \\
K^- & \bar K^0 &\displaystyle - \frac{ 2 \, \eta_8}{\sqrt 6}
\end{array}  \right) .
\end{equation}
In terms of these the lowest order Lagrangian is given by
\begin{equation}
{\cal L}_2 = \frac{F_0^2}{4} \{\langle D_\mu U^\dagger D^\mu U \rangle 
+\langle \chi^\dagger U+\chi U^\dagger \rangle \}\, ,
\end{equation}
with $D_\mu U = \partial_\mu U -i r_\mu U + i U l_\mu \,,$
and $\chi = 2 B_0 (s+ip)$ in terms of the
left and right external currents: $r(l)_\mu = v_\mu +(-) a_\mu$
and scalar and pseudo-scalar external densities: $s,p$
\cite{Gasser:1984gg}. Quark masses are included via the scalar density:
$s= {\cal M} + \cdots$. The notation 
$\langle A \rangle = Tr_F\left(A\right)$ indicates the trace over flavours.

At higher orders many more terms appear. The number of terms is listed in
Tab.~\ref{tab:LECs}. The free coefficients of those terms are called low-energy
constants (LECs). The two- and three-flavour
$p^4$ Lagrangians were constructed in \cite{Gasser:1983yg,Gasser:1984gg},
the $p^6$ Lagrangians in \cite{Bijnens:1999sh}.
\begin{table}[t!]
\begin{tabular}{ccccccc}
\hline
order  & \multicolumn{2}{c}{ 2 flavour} & \multicolumn{2}{c}{3 flavour} &
\multicolumn{2}{c|}{ PQChPT/$N_f$ flavour}\\
\hline
$p^2$ & $F,B$ & 2 & $F_0,B_0$ & 2 &  $F_0,B_0$ &  2 \\
$p^4$ & $l_i^r,h_i^r$ & 7+3 & $L_i^r,H_i^r$ & 10+2 & 
      $\hat L_i^r,\hat H_i^r$ &  11+2 \\
$p^6$ & $c_i^r$ & 52+4 & $C_i^r$ & 90+4 &  $K_i^r$ &
       112+3\\
\hline
\end{tabular}
\caption{The number of low-energy constants (LECs) at each order in the
expansion for a number of cases, the $i+j$ notation indicates the number
of mesonic + pure contact terms.}
\label{tab:LECs}
\end{table}
Including finite volume and boundary conditions does not introduce any new LECs,
other effects and interactions typically introduce (many) new LECs.

Let me just add a reminder about the main properties of ChPT:
It relates processes with different numbers of pseudo-scalars,
includes isospin and the eightfold way ($SU(3)_V$) and
unitarity and analyticity effects are included perturbatively.
The best known consequence are the chiral logarithms, e.g. for the
example of the pion mass~\cite{Gasser:1983yg}
\begin{equation}
m_\pi^2 = 2 B \hat m  + \left(\frac{2 B \hat m}{F}\right)^2
\left[ \frac{1}{32\pi^2}\mbox{\framebox{$\displaystyle\log\frac{\left(2 B \hat m\right)}{\mu^2}$}} + 2 l_3^r(\mu)\right] +\cdots
\end{equation}
with $M^2 = 2 B \hat m$ the lowest-order mass.

\section{Determination of LECs in the continuum}

One of the problems in practically using ChPT is to have values for the unknown
LECs. The original determination was done in \cite{Gasser:1983yg,Gasser:1984gg}
at the $p^4$ level. However, all needed observables are known to
order $p^6$, as reviewed in \cite{Bijnens:2006zp}.
The latest update of the LECs can be found in \cite{Bijnens:2014lea}.

The two-flavour constants, quoted in the subtraction-scale-independent
form $\bar l_i$, are
\begin{eqnarray}
\label{valueli}
\bar l_1&=&-0.4\pm 0.6\,,\qquad\qquad~~
\bar l_2 = 4.3\pm0.1\,,\qquad\qquad
\bar l_3=3.0\pm0.8\,,\qquad\qquad
\bar l_4=4.3\pm0.2\,,
\nonumber\\
\bar l_5 &=& 12.24\pm0.21\,,\qquad\qquad
\bar l_6-\bar l_5 = 3.0\pm0.3\,,\qquad
\bar l_6 = 16.0\pm0.5\pm0.7\,.
\end{eqnarray}
$\bar l_1$ and $\bar l_2$ follow from the $\pi\pi$-scattering analysis
\cite{Colangelo:2000jc}, see also \cite{Nebreda:2012ve}.
$\bar l_3$ is mainly restricted from
lattice data \cite{Aoki:2013ldr} and $\bar l_4$ from
the quark mass dependence of $F_\pi$ and the pion scalar radius
\cite{Bijnens:1998fm}. $\bar l_5-\bar l_6$ is from the pion electromagnetic
radius \cite{Bijnens:1998fm}, while $\bar l_5$ follows from
the decay $\pi\to e\nu\gamma$\cite{Bijnens:1996wm}
and $\tau$-decays \cite{GonzalezAlonso:2008rf}.

The three flavour first full $p^6$ fit was done
in \cite{Amoros:2000mc,Amoros:2001cp}. Including many more observables and new
data, a major update was done by \cite{Bijnens:2011tb}
and a final update with the same experimental input but some more information
on $p^6$ LECs in \cite{Bijnens:2014lea}.
Recent values of the weak interaction ChPT LECs can be found in
\cite{Bijnens:2004ai,Cirigliano:2011ny}. An overview of the lattice work
is the FLAG second report~\cite{Aoki:2013ldr}.

For the three-flavour case
we have that $m_K^2,m_\eta^2>> m_\pi^2$ so a question is whether ChPT
works at all in this sector. The contributions from the not very well
known $p^6$ LECs are much larger and there is the question of the importance
of $1/N_c$ suppressed terms. In \cite{Bijnens:2009zd} a large number
of observables was checked and a number of relations found that
were independent of the $p^6$ LECs and only depend on $p^4$ LECs via
loop contributions. With 76 observables we found 35 relations. For 13
of these there were enough experimental data available.
The resulting picture was that three-flavour ChPT
works but might converge slowly in some cases.

The data included for a fit of $L_1^r,\cdots,L_8^r$ are:
\begin{itemize}
\item $M_\pi,M_K,M_\eta,F_\pi,F_K/F_\pi$
\item $\langle r^2\rangle^\pi_S$, $c^\pi_S$ slope and curvature of $F_S$
\item $\pi\pi$ and $\pi K$ scattering lengths  $a^0_0$, $a^2_0$, $a_0^{1/2}$
and $a_0^{3/2}$. 
\item Value and slope of $F$ and $G$ in $K_{\ell4}$
\item $\frac{m_s}{\hat m}=27.5$ (lattice)
\item $\bar l_1,\ldots,\bar l_4$
\end{itemize}
This corresponds to $17+3$ inputs and we have
8 $L_i^r$ and 34 combinations of $C_i^r$ to fit, a clearly ill-defined problem.
The older fits \cite{Amoros:2001cp} (ABC01), \cite{Bijnens:2011tb}(BJ12)
used a simple
resonance estimate of the $C_i^r$, this was complemented by more
input on the $C_i^r$ from other models and various estimates
and a requirement of not too large $p^6$ corrections
the meson masses in \cite{Bijnens:2014lea} (BE14). The resulting values
of the fits are shown in Tab.~\ref{tab:BE14}.
\begin{table}[t!]
\begin{tabular}{l r r r r}
\hline
             & ABC01 &  JJ12          &  $L_4^r$ free    & BE14            \\
\hline                                                         
\rule{0cm}{0cm}& old data   &                &                  &                 \\
$10^3 L_1^r$ &$0.39(12)$ &$0.88(09)$   &$0.64(06)$     & $0.53(06)$     \\
$10^3 L_2^r$ &$0.73(12)$  &$0.61(20)$  &$0.59(04)$    & $0.81(04)$    \\
$10^3 L_3^r$ &$-2.34(37)$ &$-3.04(43)$ &$-2.80(20)$   & $-3.07(20)$   \\
$10^3 L_4^r$ &$\equiv0$      &$0.75(75)$  &$0.76(18)$    & $\equiv0.3$      \\
$10^3 L_5^r$ &$0.97(11)$ &$0.58(13)$   &$0.50(07)$     & $1.01(06)$     \\
$10^3 L_6^r$ &$\equiv 0$    &  $0.29(8)$  &   $0.49(25)$  & $0.14(05)$    \\
$10^3 L_7^r$ &$-0.30(15$&$-0.11(15)$  &$-0.19(08)$    & $-0.34(09)$   \\
$10^3 L_8^r$ &$0.60(20)$ &$0.18(18)$   &$0.17(11)$     & $0.47(10)$ \\
 \hline                                                        
$\chi^2$     & $0.26$       &$1.28$          &$0.48$            & $1.04$   \\ 
dof          & 1            & 4              & ?                & ?\\
$F_0$ [MeV]  & 87           & 65             & 64               & 71\\
\hline
\end{tabular}
\caption{Values of the $p^4$ three-flavour ChPT LECs, $L_i^r$ in the major fits
performed at two-loop order. $dof$ stands for degrees of freedom.}
\label{tab:BE14}
\end{table}

Many prejudices, as described in detail in \cite{Bijnens:2014lea}, were
used in fixing the values of the $C_i^r$. The final values chosen are
all ``reasonable'' and compatible with existing determinations.
The large $N_c$ suppressed constant $L_4^r$, if left free, is rather large.
We therefore restricted it to the expected range. Surprisingly, this lead
to the values of $L_6^r$ and $2L_1^r-L_2^r$ also being small and compatible
with large $N_c$ arguments. The values for the $L_i^r$ are compatible
with existing lattice determinations as well. The convergence is reasonable,
but enforced for the masses, as can be seen from the examples:
\begin{eqnarray}
\mbox{Mass:}\qquad\qquad~~
m^2_\pi/m^2_{\pi phys}  &=&1.055(p^2)-0.005(p^4)-0.050(p^6)\,, \nonumber\\
m^2_K/m^2_{K phys}      &=&1.112(p^2)-0.069(p^4)-0.043(p^6)\,,\nonumber\\
m^2_\eta/m^2_{\eta phys}&=&1.197(p^2)-0.214(p^4)+0.017(p^6)\,,\nonumber\\
\mbox{Decay constants:}\qquad
{F_\pi}/{F_0} &=& 1.000(p^2)+0.208(p^4)+0.088(p^6)\,,\nonumber\\
{F_K}/{F_\pi} &=& 1.000(p^2)+0.176(p^4)+0.023(p^6)\,. \nonumber\\
\mbox{Scattering:}\qquad\qquad\qquad
{a^0_0} &=& 0.160(p^2)+0.044(p^4)+0.012(p^6)\,,\nonumber\\
a^{1/2}_0 &=& 0.142(p^2)+0.031(p^4)+0.051(p^6)\,.
\end{eqnarray}

\section{Finite volume}

An example of extra effects that can be included is the use of ChPT to
study the effects of a finite volume. Finite volume effects were studied
first in a general way by L\"uscher \cite{Luscher:1985dn} and
soon introduced in ChPT by Gasser and
Leutwyler~\cite{Gasser:1986vb,Gasser:1987zq}.
In particular, \cite{Gasser:1987zq} proved that no new LECs were needed.
They calculated $m_\pi, F_\pi$ and $\langle\bar q q\rangle$ to one-loop
in the equal mass case. Note that the remainder will be in the
$p$-regime with $m_\pi L \ge 1$. $L$ is the size of the finite volume.
ChPT will be useful since the convergence
is given by the rho mass with $1/m_\rho\approx 0.25$~fm, while the
finite volume effects are controlled by $1/m_\pi\approx1.4$~fm. It will
often be needed to go beyond the leading $\exp(-m_\pi L)$ behaviour.
An introduction and more references can be found in \cite{Golterman:2009kw}.

A partial overview of existing results at finite volume is:
Masses and decay constants for three flavours at one-loop
\cite{Becirevic:2003wk,DescotesGenon:2004iu,Colangelo:2005gd},
$m_\pi$ at two-loop order in two-flavour ChPT \cite{Colangelo:2006mp}
and the quark-anti-quark vacuum-expectation value at two-loops
in three-flavour ChPT \cite{Bijnens:2006ve}.
Other examples are including a twisted mass \cite{Colangelo:2010cu}
and twisted boundary conditions \cite{Sachrajda:2004mi} in ChPT.
I will now concentrate on two recent developments.

\subsection{Twisted boundary conditions}

On a lattice with a finite size given by $L$, components of spatial momenta
are restricted by $p^i= 2\pi n^i/L$ with $n^i$ integer. That means that in
practice very few low momenta are available.
One way to allow for more momenta is to put a boundary condition
on some of the quark fields in some directions via
$q(x^i+L) = e^{i\theta^i_q}q(x^i)$.
Then allowed momenta are $p^i = \theta^i/L + 2\pi n^i/L$.
Varying the $\theta^i_q$ allows to map out momentum space on
the lattice much better \cite{Bedaque:2004kc}. The finite box
breaks rotational symmetry down to cubic symmetry but twisting
reduces it even further. Consequences are:
\begin{itemize}
\item $m^2(\vec p) = E^2-\vec p^2$ is not constant.
\item There are typically more form-factors than in infinite volume.
\item In general quantities can depend on many more components of the momenta,
not just Lorentz-invariant products.
\item Charge conjugation involves a change in momentum.
\item The boundary conditions can break isospin.
\end{itemize}
\begin{figure}[t!]
\begin{minipage}{0.49\textwidth}
\includegraphics[width=\textwidth]{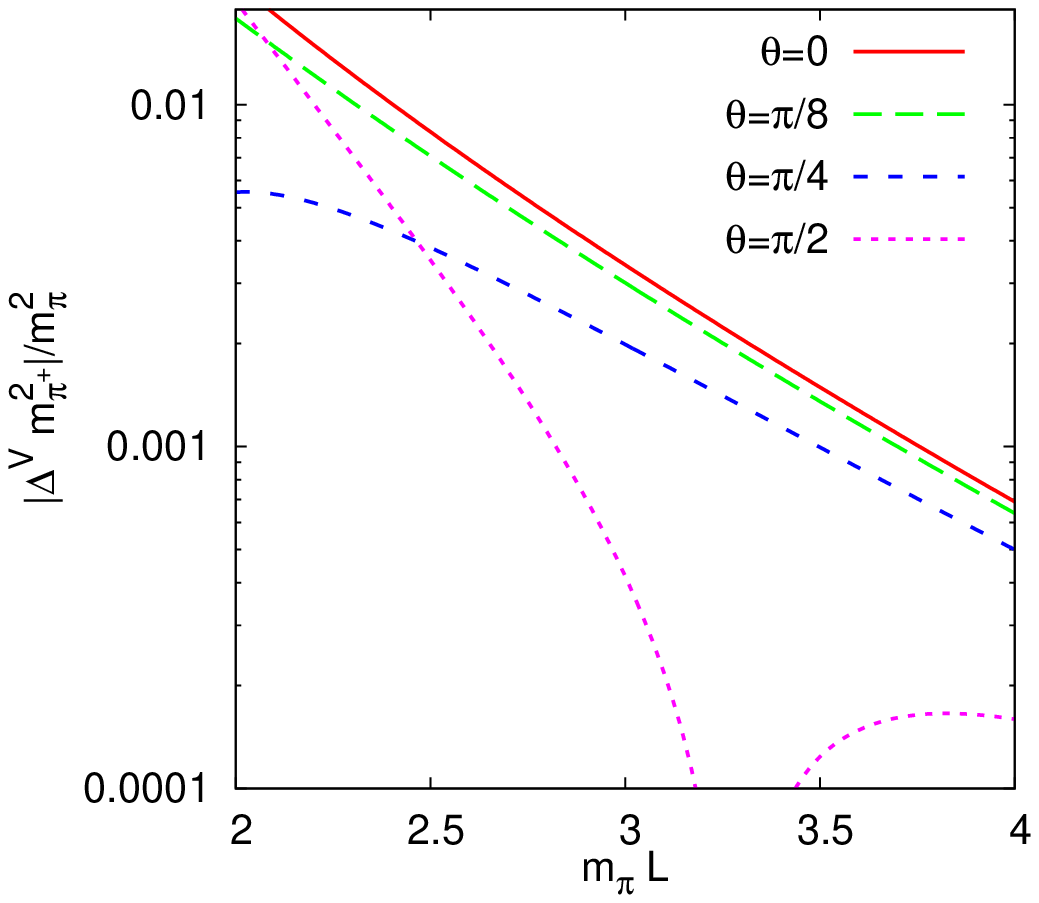}
\centerline{(a)}
\end{minipage}
\begin{minipage}{0.49\textwidth}
\includegraphics[width=\textwidth]{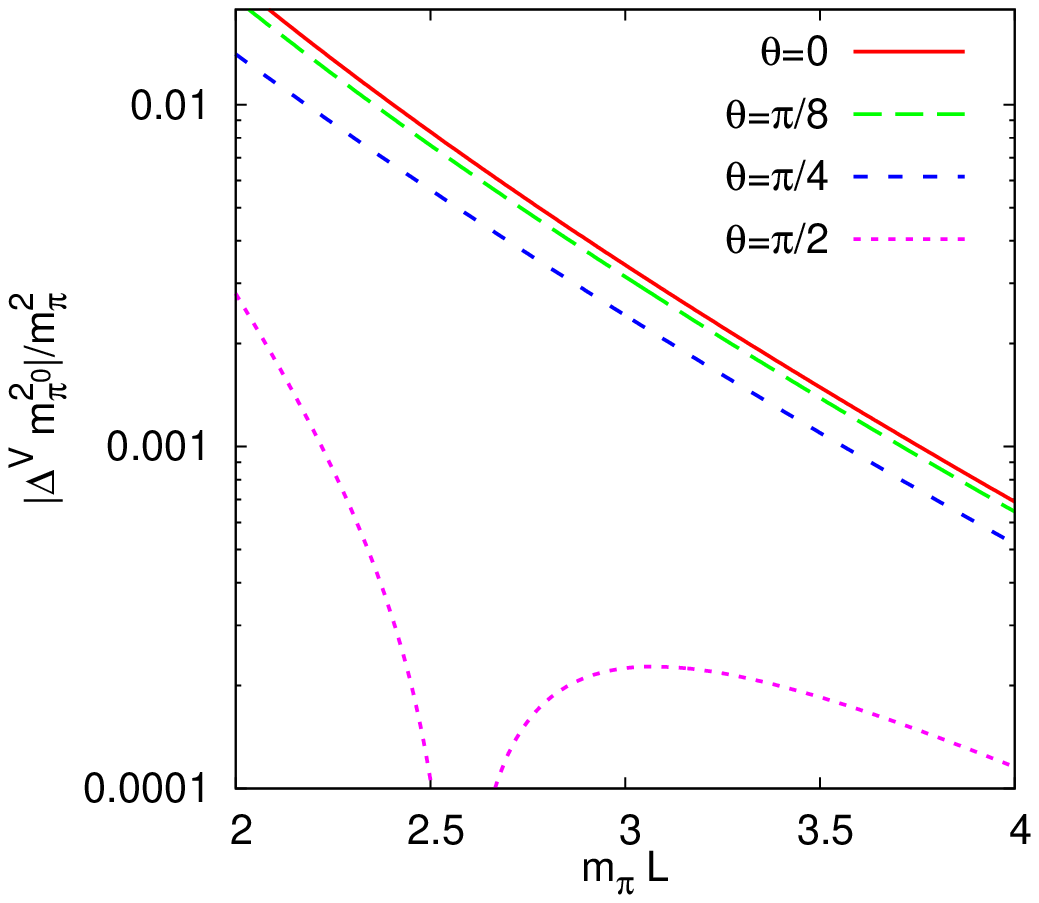}
\centerline{(b)}
\end{minipage}
\caption{Finite volume corrections to the charged and neutral pion
mass squared as a function of $m_\pi L$ for several values of the up quark
twist angle $\theta$. (a) Charged pion (b) neutral pion.
Plots from  \cite{Bijnens:2014yya}.}
\label{figmpitwisted}
\end{figure}
As a first example I show the finite volume corrections to the 
charged and neutral pion mass \cite{Bijnens:2014yya} in
Fig.~\ref{figmpitwisted}. The plots show the finite volume correction
$\Delta^V m^2 = m^{2V}-m^{2\,V=\infty}$
as a function of $m_\pi L$ for several values of $\theta$ with
$\vec\theta_u=(\theta,0,0)$ and $\vec\theta_d=\vec\theta_s=0$.
The relation with the earlier work in \cite{Sachrajda:2004mi}
and \cite{Jiang:2006gna} is discussed in detail in \cite{Bijnens:2014yya}.
Note that the finite volume correction is very dependent on the twist-angle.

The matrix-element for the decay constant has extra terms
\begin{equation}
\left< 0 |A_\mu^M | M(p) \right> = i\sqrt{2}F_M p_\mu + i\sqrt{2}F^V_{M\mu}\,.
\end{equation}
These are required such that the Ward identities are
satisfied \cite{Bijnens:2014yya} and the extra components can be quite sizable.

The vector-form factors also require extra components
\cite{Bijnens:2014yya}:
\begin{equation} 
\left<M^\prime(p^\prime)|j_\mu|M(p)\right> = f_ \mu
 = f_{+} (p_\mu+p_\mu^\prime) 
    + f_{  -} q_\mu
    +  h_{\mu}\,.
\end{equation}
 earlier work on two flavours is \cite{Jiang:2006gna}. Note that
the vector current satisfies the Ward identities, contrary to what is sometimes
stated but
$ q^\mu f_\mu = (p^2-p^{\prime2}) f_+ + q^2 f_- + q^\mu h_\mu =0$ requires
to include all components and the use of the correct finite volume
masses for $p^2$ and $p^{\prime2}$. 

\begin{figure}[t!]
\begin{minipage}{0.49\textwidth}
\includegraphics[width=\textwidth]{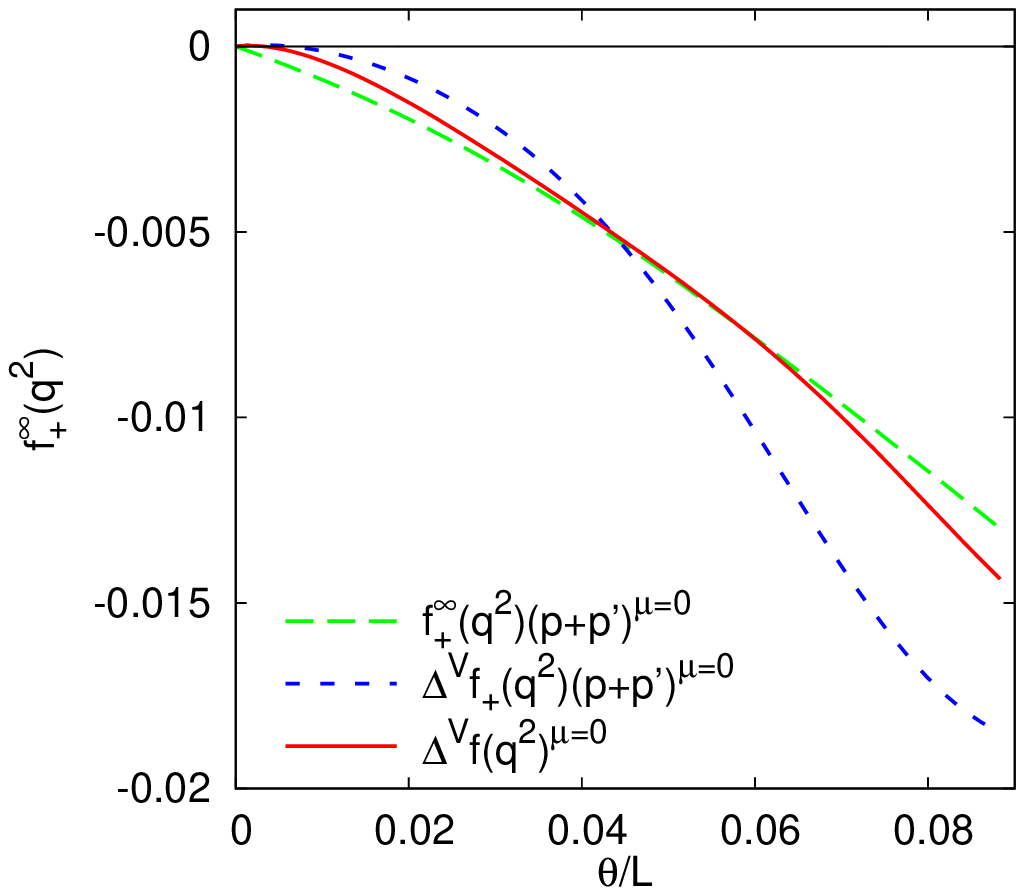}
\centerline{(a)}
\end{minipage}
\begin{minipage}{0.49\textwidth}
\includegraphics[width=\textwidth]{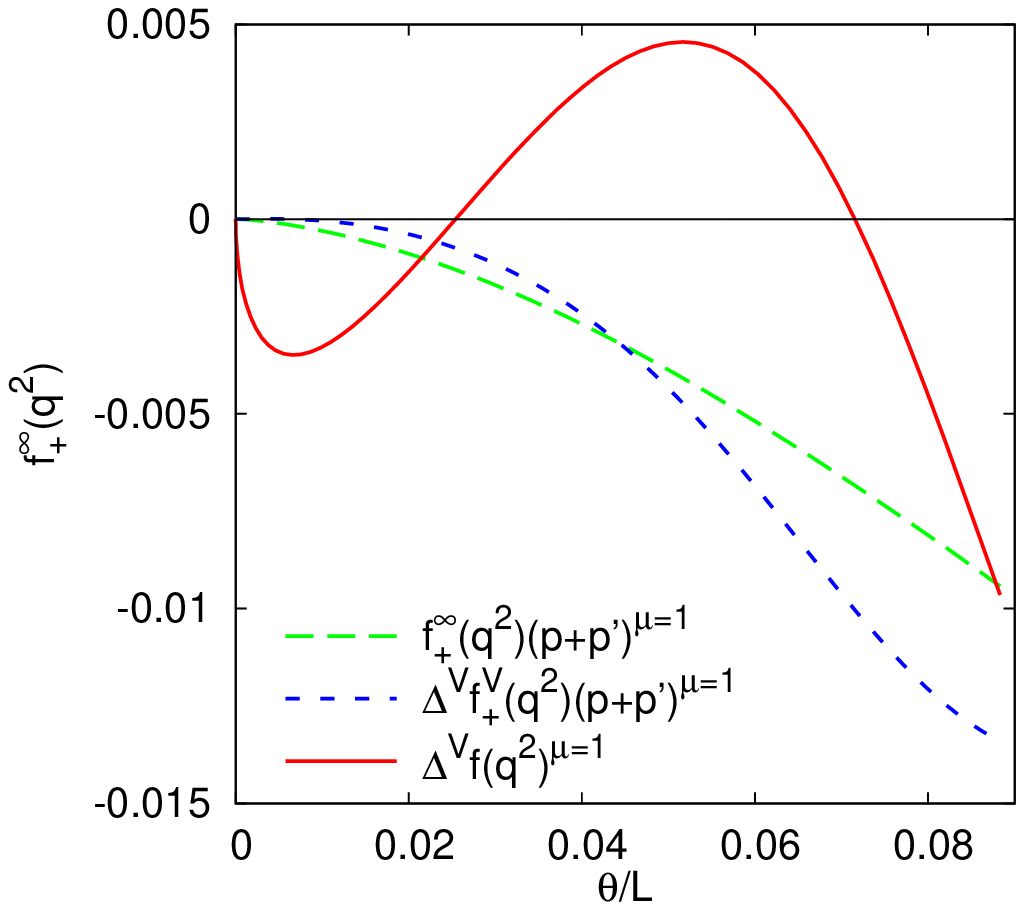}
\centerline{(b)}
\end{minipage}
\caption{Finite volume corrections and infinite volume one-loop
part of the components of the $\pi^+$-$\pi^0$ vector transition form-factors.
(a) $\mu=t$ component (b) $\mu=x$ component.
Plots from  \cite{Bijnens:2014yya}.}
\label{fig:FV}
\end{figure}

The lattice determination of the pion electromagnetic form-factor
from the $\pi^+$-$\pi^0$ transition amplitude
\begin{equation}
f_\mu = -\frac{1}{\sqrt{2}} \langle\pi^0(p^\prime)|\bar d \gamma_\mu u|\pi^+(p)\rangle
 = \left(1+f^{\infty}_++\Delta^V f_+ \right)(p+p^\prime)_\mu
   +\Delta^V f_- q_\mu + \Delta^V h_\mu
\label{deffmu}
 \end{equation}
requires all the finite volume corrections.
In Fig.~\ref{fig:FV} the corrections needed are shown for
the time and $x$ spatial component of the form-factor $f_\mu$ of
(\ref{deffmu}).
Plotted is also for comparison the pure one-loop contribution to the
infinite volume form-factor $f_+^\infty$.

\subsection{Masses at two-loops} 

\begin{figure}[t!]
\begin{minipage}{0.49\textwidth}
\includegraphics[width=\textwidth]{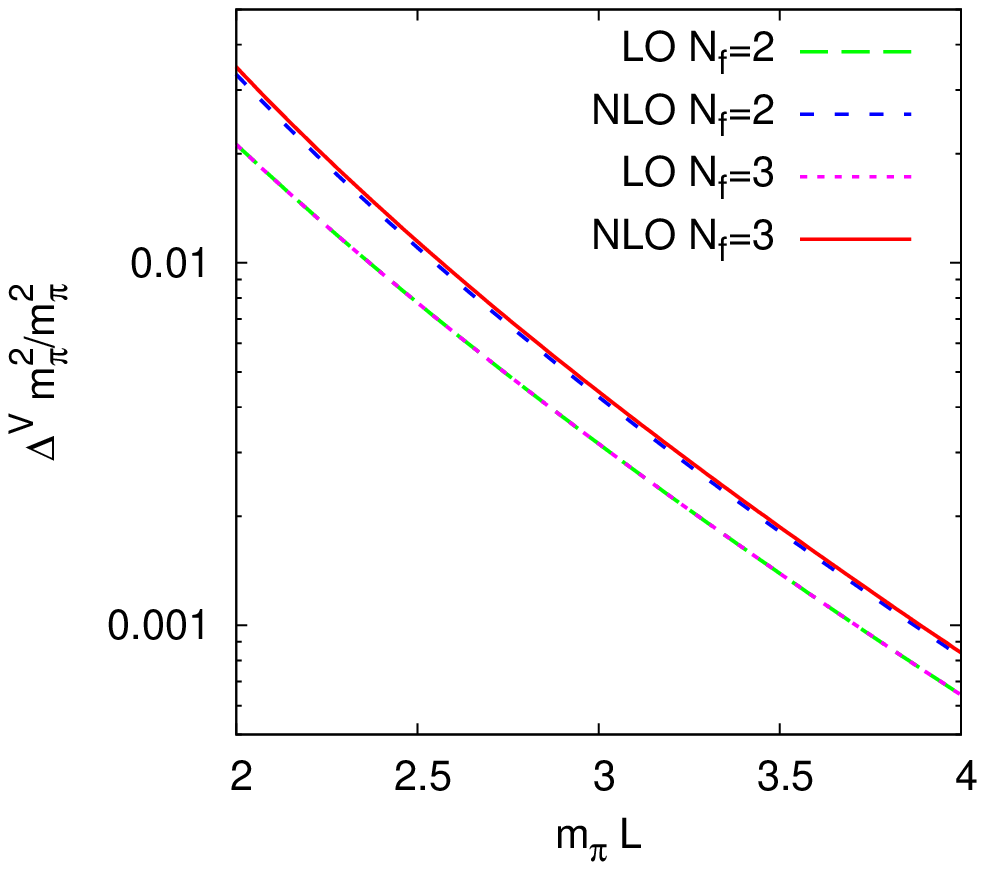}
\centerline{(a)}
\end{minipage}
\begin{minipage}{0.49\textwidth}
\includegraphics[width=\textwidth]{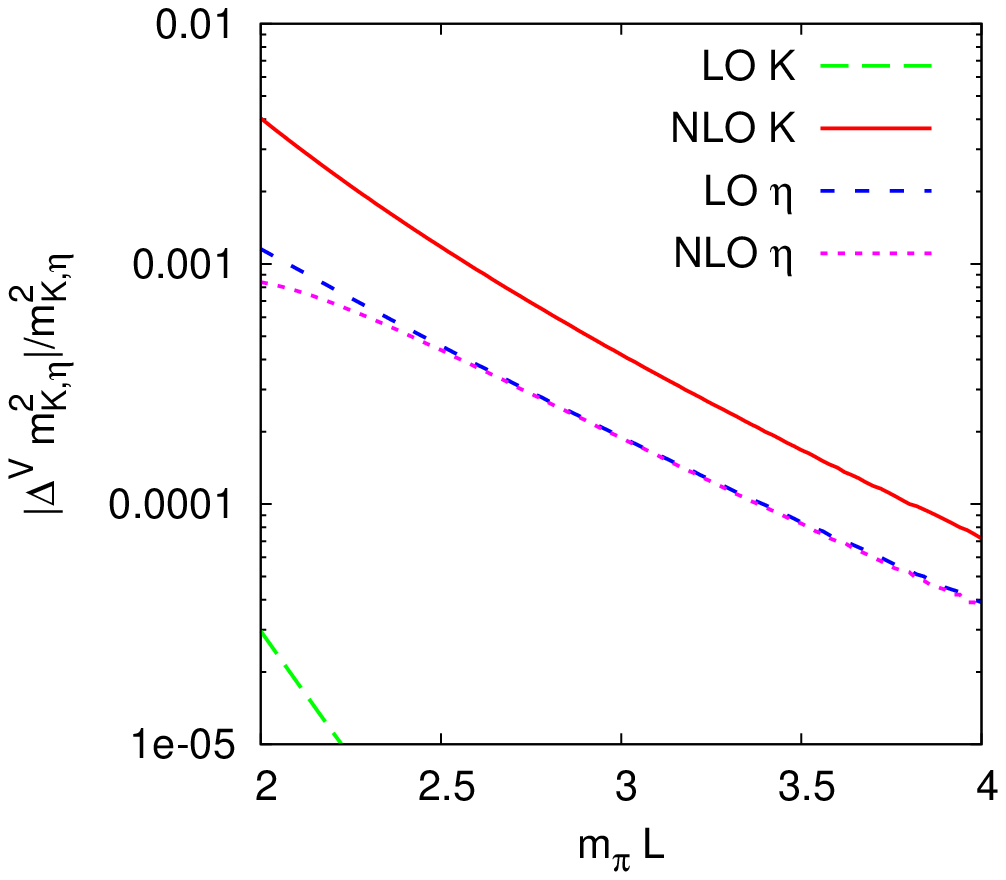}
\centerline{(b)}
\end{minipage}
\caption{Finite volume corrections at two-loop order to the meson masses
squared. Shown is the relative correction $\Delta^V m_\pi^2/m_\pi^2
 = m_\pi^{V2}/m_\pi^{V=\infty 2}-1$. LO is the $p^4$ or one-loop results.
NLO is the $p^6$ or two-loop result.
(a) Pion mass in two- and three-flavour ChPT (b) Kaon and eta mass.}
\label{figmassFV}
\end{figure}

The finite volume correction for the meson masses and decay constants in
three-flavour ChPT is in progress \cite{BR}. As was already visible in
the two-flour two-loop calculation of \cite{Colangelo:2006mp}, the main
obstacle for a full two-loop calculation is the finite volume sunset
integrals. These were derived for the most general mass case in
\cite{Bijnens:2013doa}, thus paving the way for a full two-loop evaluation.
Some preliminary results are shown in Fig.~\ref{figmassFV}.
For the pion mass at order $p^4$ the two- and three-flavour result differ
by kaon and eta loops. These are numerically very small. The $p^6$ results
are also in good agreement with each other. The kaon mass
at order $p^4$ has only a very small correction since there is no pion-loop
contribution. The $p^6$ contribution is of the expected size.
For the $\eta$, there is a cancellation between the pure two-loop contribution
and the $L_i^r$-dependent part at $p^6$ resulting in a very small correction
in total. The $p^4$ contribution is suppressed by an extra factor of
$m_\pi^2/m_\eta^2$. More results and details will be published in \cite{BR}.

\section{Beyond QCD}

There are other symmetry breaking patterns possible in generic
gauge theories. Early examples are related to technicolour
\cite{Peskin,Preskill,Dimopoulos} and some might be useful to
avoid the sign problem in high density QCD lattice simulations \cite{Kogut}.
The equal mass case requires for many quantities the same integrals as
needed for $\pi\pi$-scattering in two-flavour
ChPT~\cite{Bijnens:1997vq,Bijnens:1995yn}. The lattice studies of these
type of theories was discussed in the talk by E.~Pallante.
One often wants to extrapolate to zero fermion masses from the lattice data
here. ChPT can help there, just as for the QCD case. A number of quantities
were studied for $N$ equal mass flavours for the complex, real and
pseudo-real case \cite{Bijnens:2009qm,Bijnens:2011fm,Bijnens:2011xt}
at two-loop order.
References to earlier one-loop work can be found in our work and \cite{Kogut}.

Generically the fermions can be in a complex, real or pseudo-real representation
of the gauge group. Examples are of the first case QCD, the second case
any group with fermions in the adjoint representation and the last case
an $SU(2)$ gauge group with fermions in the fundamental representation.
In the latter two cases anti-quarks are in the same representation
as the quarks leading to larger global chiral symmetry group. Assuming
that a condensate forms similar to QCD, we get the breaking patterns:
\begin{itemize}
\item $SU(N)\times SU(N)/SU(N)$ (complex)
\item $SU(2N)/SO(2N)$ (real)
\item $SU(2N)/Sp(2N)$ (pseudo-real)
\end{itemize}
The three cases can be dealt with in very similar fashion.

The standard
QCD case has a vector $q^T = (q_1\cdots q_{N_F})$ and the chiral symmetry
transformation under $(g_L,g_R)\in G=SU(N_F)_L\times SU(N_F)_R$
is $q_L\to g_L q_L, q_R\to g_R q_R$. The condensate 
 $\langle\overline q_{Lj} q_{Ri}\rangle= -v\Sigma_{ij}$
is described by a unitary matrix $\Sigma$. The vacuum
expectation value is $\langle\Sigma\rangle=1$, the unity matrix, such that
for $g_L=g_r$ the vacuum remains invariant under
$\Sigma\to g_R\Sigma g_L^\dagger$. The conserved symmetry group is thus
$H=SU(N_F)_V$.

The case with $N_F$ fermions in a real representation of the gauge group
can be described by a $2N_F$ vector
$\hat q^T = (q_{R1}~\ldots~q_{R N_F}~\tilde q_{R1}~\ldots~\tilde q_{R N_F})$
with $\tilde q_{Ri} \equiv C \bar q^T_{Li}$. $C$ is charge conjugation.
The global chiral symmetry is thus $G=SU(2N_F)$ with $\hat q\to g\hat q$.
The vacuum expectation value  $\langle\overline q_j q_i\rangle$ is really
\begin{equation}
\Sigma_{ji}=\langle (\hat q_j)^T C \hat q_i\rangle \propto J_{Sij}
\qquad \mathrm{with} \qquad 
J_S=
\left(\begin{array}{cc} 0 & I\\
I & 0\end{array}\right)\,.
\end{equation}
$\Sigma$ is $2N_F\times 2N_F$ and $\Sigma\to g\Sigma g^T$ for $g\in G$.
The vacuum is conserved if
 $g J_S g^T = J_S$ $\Longrightarrow$. The conserved part is $H=SO(2N_F)$.

For $N_F$ fermions in a pseudo-real representation the situation is similar
but  $\hat q^T = (q_{R1}~\ldots~q_{R N_F}~\tilde q_{R1}~\ldots~\tilde q_{R N_F})$
with $\tilde q_{R\alpha i} \equiv \epsilon_{\alpha\beta}C \bar q^T_{L\beta i}$
instead. $q_{R i}$ transforms under the gauge group as $q_{R\alpha i}$.
The global chiral symmetry is thus again $G=SU(2N_F)$ with $\hat q\to g\hat q$.
The vacuum expectation value  $\langle\overline q_j q_i\rangle$ corresponds now
to
\begin{equation}
-v \Sigma_{ji}=\epsilon_{\alpha\beta}\langle (\hat q_{\alpha j})^T C \hat q_{\beta i}\rangle \propto J_{Aij}
\qquad\mathrm{with}\qquad 
J_A=
\left(\begin{array}{cc} 0 & -I\\
I & 0\end{array}\right)\,.
\end{equation}
$\Sigma$ is again a $2N_F\times 2N_F$ matrix
and $\Sigma\to g\Sigma g^T$ for $g\in G$.
The vacuum is conserved if
 $g J_A g^T = J_A$ with a conserved global symmetry $H=Sp(2N_F)$.

ChPT for the three cases is extremely similar if we define
$u=\exp(i \phi^a X^a/(\sqrt2F))$ \cite{Bijnens:2009qm} with the
$X^a$ the generators of an $SU(N_F)$ (complex case),
or of an $SU(2N_F)$ satisfying $X^a J_S = J_S X^{aT}$ (real)
or  $X^a J_A = J_A X^{aT}$ (pseudo-real). Note that these are not the usual
ways of parametrizing $Sp(2N_F)$ or $SO(2N_F)$ matrices but related.
As a consequence the Lagrangians constructed for the $N_F$ flavor complex
case \cite{Bijnens:1999sh} can be taken, but might not be minimal.
Also the divergence structure for the complex case is known
\cite{Bijnens:1999hw}, providing a check on the calculations.

\begin{figure}
\begin{minipage}{0.49\textwidth}
\includegraphics[width=0.99\textwidth]{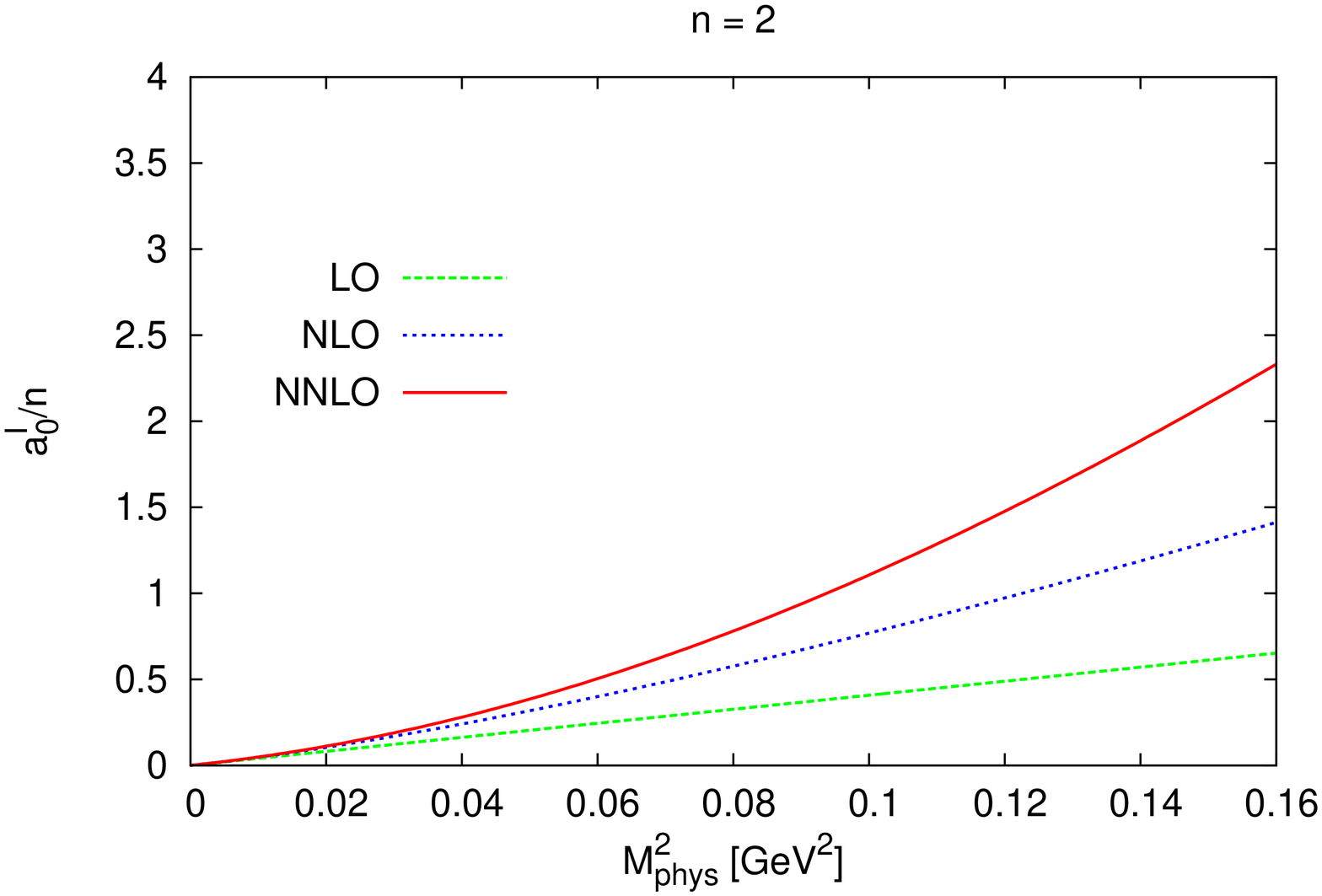}
\includegraphics[width=0.99\textwidth]{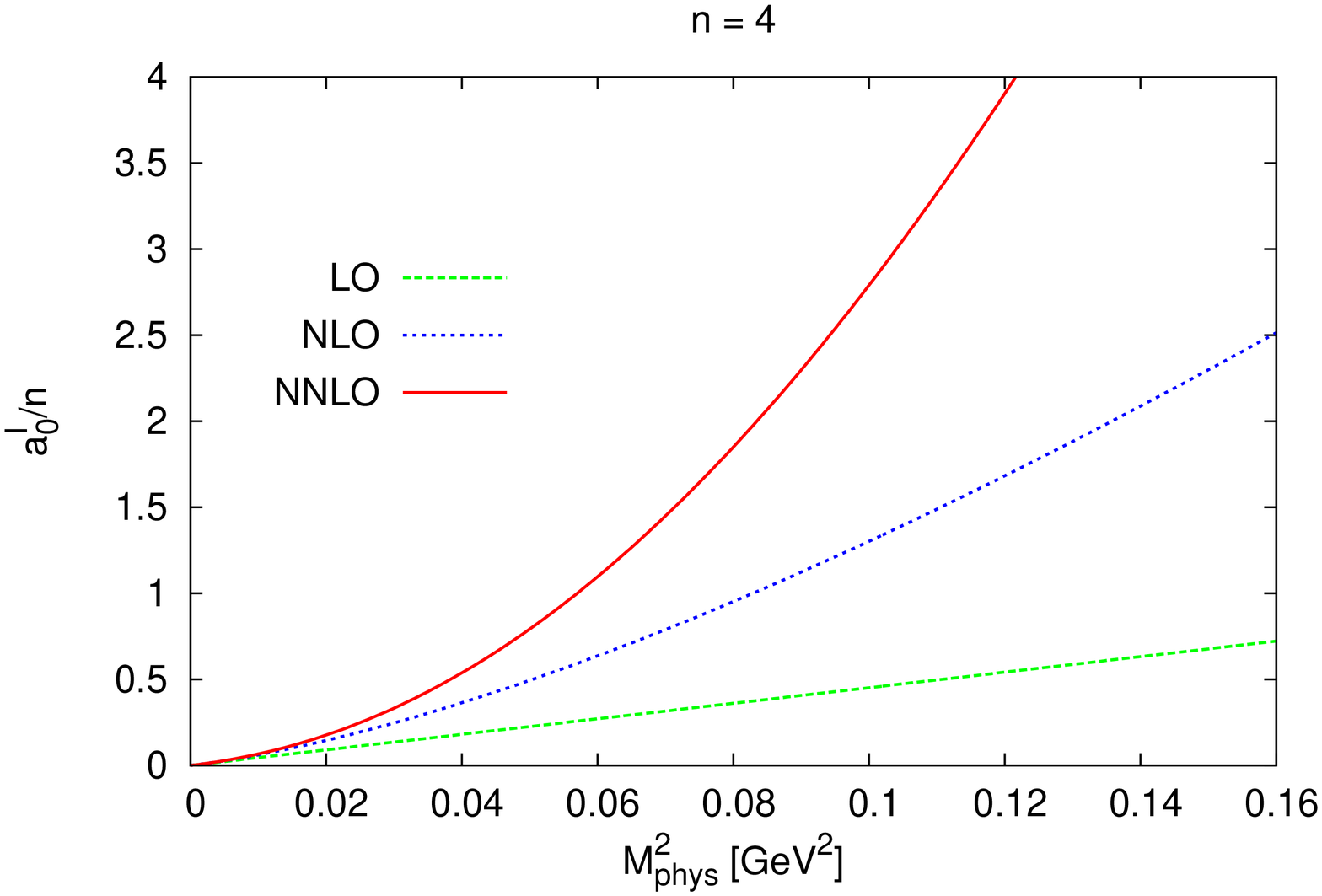}
\end{minipage}
\begin{minipage}{0.49\textwidth}
\includegraphics[width=0.99\textwidth]{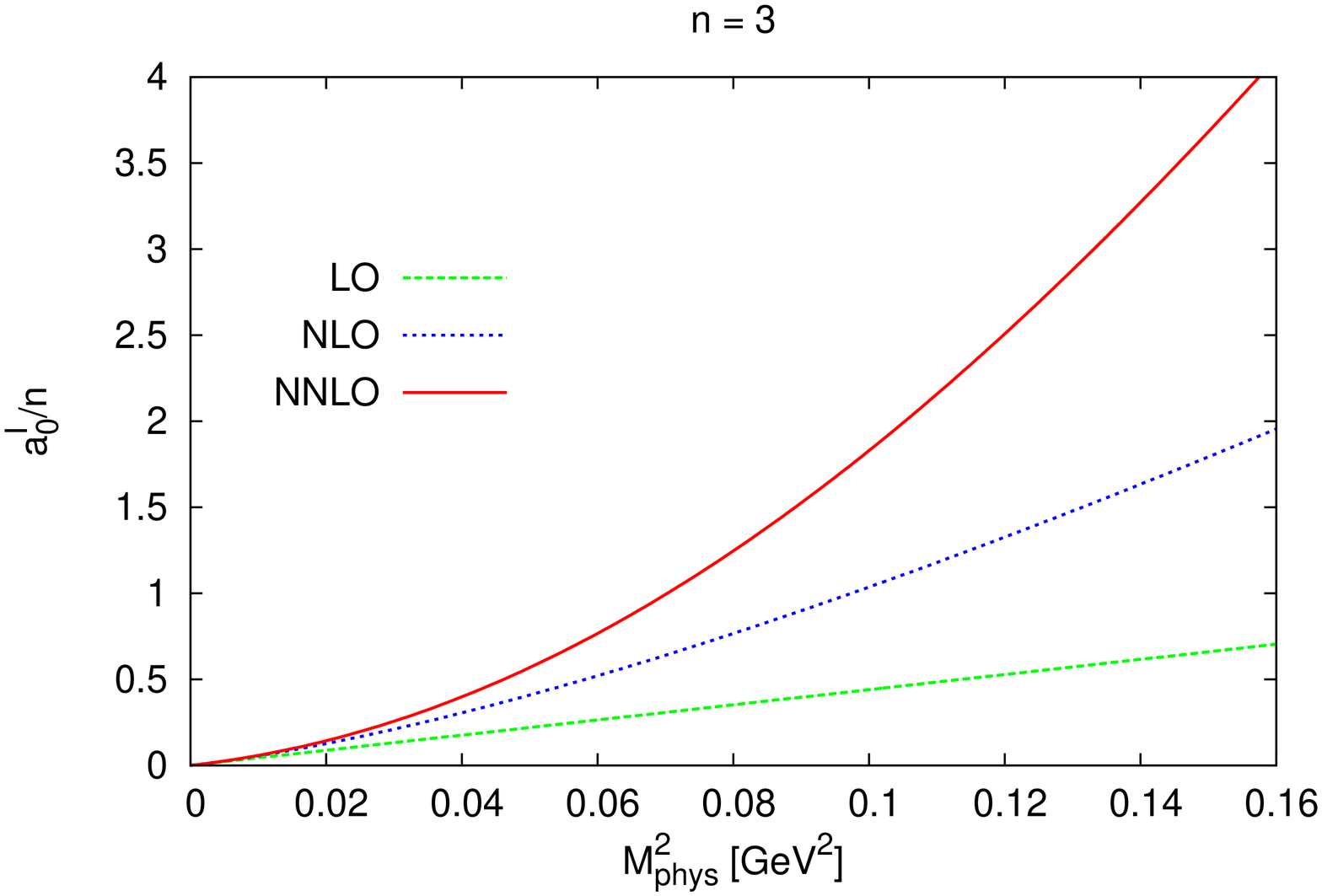}
\includegraphics[width=0.99\textwidth]{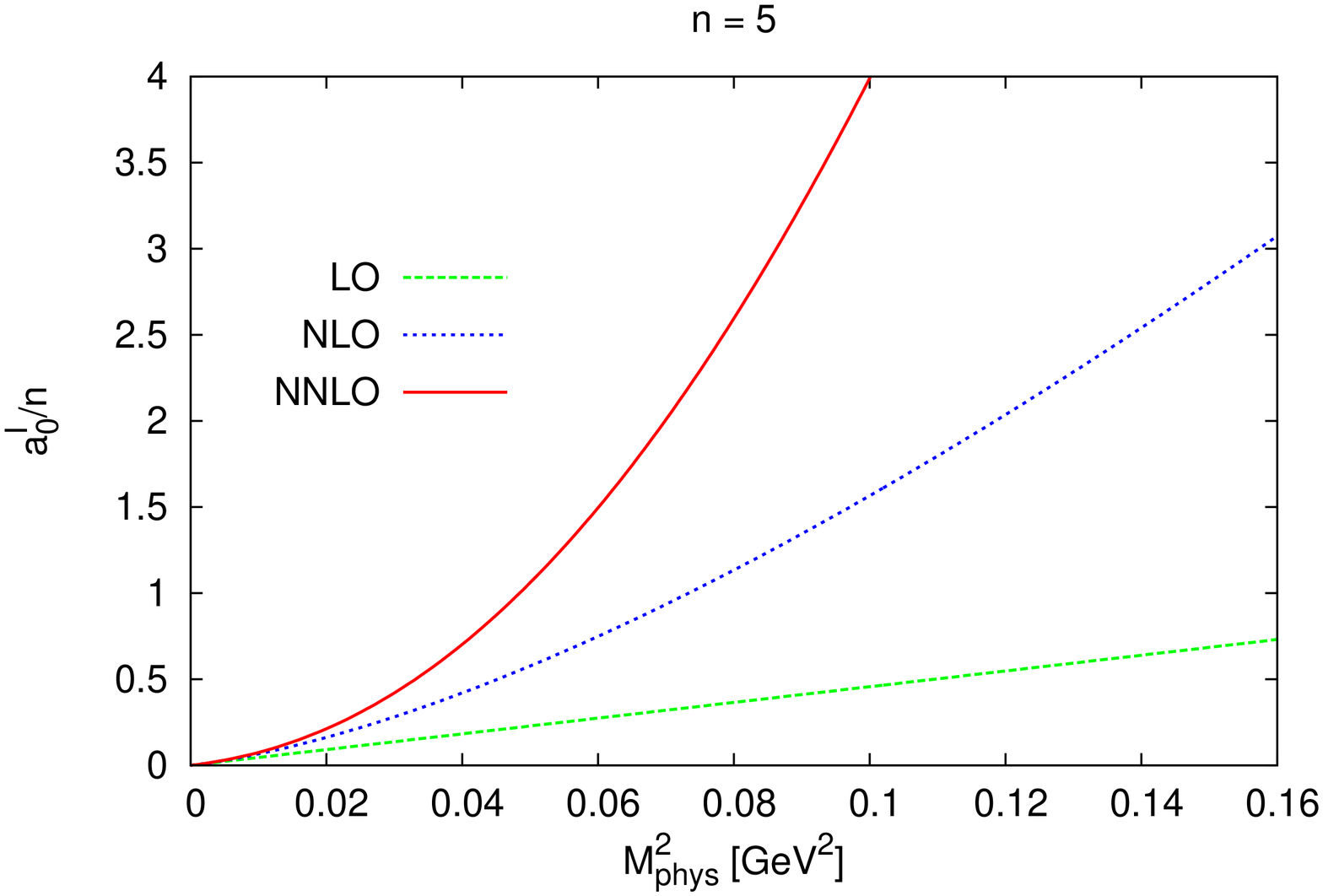}
\end{minipage}
\caption{The singlet scattering length as a function of the meson mass. The scale is set by the decay constant $F\approx F_\pi$.
Plots from \cite{Bijnens:2011fm}. For technicolour applications the mass
and decay constant should be scaled up accordingly.}
\label{figMM}
\end{figure}

The expressions for masses, decay constants and vacuum expectation values
to two-loop order can be found in \cite{Bijnens:2009qm} and are known fully
analytically. The meson-scattering case can be written in terms of two
amplitudes $B(s,t,u)$ and $C(s,t,u)$ \cite{Bijnens:2011fm},
analoguous to $A(s,t,u)$
defined in $\pi\pi$-scattering, see e.g. \cite{Bijnens:1997vq}.
The possible intermediate states are a little more complicated than for
$\pi\pi$-scattering. All scattering formulas are fully analytically obtained
in \cite{Bijnens:2011fm}. For explicit expressions I refer to that paper.
As an example, I show the single meson-scattering length as a function of
$n=N_F$ for the complex case in Fig.~\ref{figMM}.

The last application we did was to perform the calculations necessary to
extract the $S$-parameter \cite{Bijnens:2011xt}.
 
\section{Leading logarithms}

The last application is the calculation of leading logarithms (LL) in EFT and
especially mesonic ChPT. Leading logarithms are the following, take as an
example an observable quantity $F$ dependent on a single physical scale $M$.
The dependence on the subtraction scale $\mu$ in field theory is typically
logarithmic:
\begin{equation}
 F= F_0 + F_1^1 { L} + F^1_0 + F_2^2{ L^2} + F^2_1 L + F^2_0 
+ F_3^3{  L^3}
 +\cdots
\,\qquad L=\log\left(\mu/M\right)\,.
\end{equation}
The coefficients $F^i_j$ are $i$ loop-level and $j$ logarithm-level.
The terms with $F^m_m$ are called the leading logarithm terms.
These terms are easier to calculate than the remaining ones at the same
loop level. The underlying reason is that physical quantities must be
independent of the subtraction scale, $ \mu\left( dF/d\mu\right)\equiv 0$
and that divergences in local quantum field theory are always local.

In a renormalizable quantum field theory the leading logarithms can be
calculated by a simple one-loop calculation using the renormalization group.
In an EFT this is not quite so simple since at each order in the expansion
new terms in the Lagrangian occur. Weinberg \cite{Weinberg:1978kz}
showed that the leading logarithms at two-loop level could be obtained
from one-loop calculations only. The full two-loop leading logarithm was
calculated with these Weinberg consistency conditions in \cite{Bijnens:1998yu}.
This was expected to work similarly to all orders and proven to do so
in \cite{Buchler:2003vw}, an alternative diagrammatic proof is in
\cite{Bijnens:2009zi}. The underlying argument is that
at $n$-loop order, ($\hbar^n$), all the divergences must cancel.
For $d=4-w$ all terms of the form $1/w^i \log^j\mu$
with $i=1,\ldots,n$ and $j=0,\ldots,n-1$ must cancel.
For the leading logarithms the $n$ conditions with $i+j=n$ give a sufficient
amount of relations that the leading logarithms can be obtained from one-loop
diagrams only, the conditions with $i+j=n-1$ show that for the next-to-leading
logarithms two-loop diagrams are required and so on.

\begin{figure}[t!]
\includegraphics[width=0.7\textwidth]{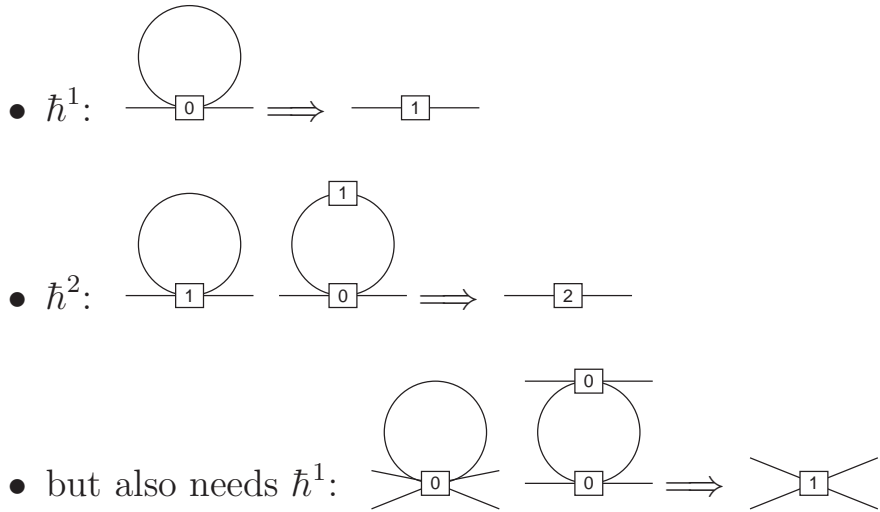}
\caption{The reason for the increase in complexity with the loop-order for
leading logarithms in EFT. The index in the vertices shows the loop-order of
the vertex needed. Top line: at one-loop we need the one-loop diagram for the
and it gives us the mass one-loop counter-term and logarithm.
Middle line: at two-loop order we need the two one-loop diagrams to get the
two-loop mass counter-term and leading logarithm. It needs the one-loop
scattering counter-term as well. Bottom line: the extra one-loop diagrams needed
to get one-loop scattering counter-term.}
\label{figRGE}
\end{figure}

The problem is that each order new terms in the Lagrangians show up, so new
one-loop diagrams are required and new vertices at each new order.
The problem is illustrated in Fig.~\ref{figRGE} for the case of the mass
at two-loop order. We need in general both new vertices of higher order but also
new vertices with more external legs.

Actually, in the massless case matters simplify somewhat, no vertices with
more external legs are needed than already appear at one-loop. The reason is
that massless tadpoles vanish.
This was used first in \cite{Bissegger:2006ix} for the scalar two-point
function LL to five loops. In
\cite{Kivel:2008mf,Kivel:2009az,Koschinski:2010mr,Polyakov:2010pt} 
a clever Legendre polynomial parametrization of the meson-meson scattering
vertices allowed to obtain the divergences at all orders via
a recursion relation that in some limits can even be solved analytically.
Large $N$ in the sigma model agreed with the older work, see e.g.
\cite{Coleman:1974jh}. 
Treated were meson-meson scattering, scalar and vector form-factors.
It was typically found that large $N$ is not a
good approximation.

We realized in \cite{Bijnens:2009zi} that a construction of a minimal
Lagrangian at each order is not necessary. When calculating the
divergences using a method that preserves the underlying symmetry the produced
divergence structure will automatically have the correct symmetry and
reducing it to its most minimal form or even rewriting it in a fully symmetric
form is not needed. The consequence is that things can be computerized
and simply let run using {\sc FORM} \cite{FORM}. We first pushed the massive
nonlinear $O(N)$ model to rather large orders for the masses, form-factors
and scattering in \cite{Bijnens:2009zi,Bijnens:2010xg} and solved the large
$N$-limit to all orders also for the massive case using gap equation techniques.
A very strong check of the result is to use different parametrizations
of the lowest-order Lagrangian. This should give the same results in the end
but intermediate expressions are very different.

An example result is the mass to fifth order via
\begin{equation}
M^2_{\mathrm{phys}}= M^2(1+a_1 L_M+a_2L^2_M+a_3 L^3_M+...)
\end{equation}
The coefficients $a_i$ are shown in Tab.~\ref{tab:ai} to five loops.
\begin{table}
\begin{tabular}{c c l}
\hline
i & $a_i$, $N=3$ & $a_i$ for general $N$\\
\hline
\hline
\rule{0cm}{4mm}1 & $-\frac{1}{2}$        & $ 1 - \frac{N}{2}$\\[1mm]
\hline
\rule{0cm}{4mm}2 & $\frac{17}{8}$       & $\frac{7}{4}
          - \frac{7 N}{4}
          + \frac{5~N^2}{8}$\\[1mm]
\hline
\rule{0cm}{4mm}3 & $-\frac{103}{24}$     & $ \frac{37}{12}
          - \frac{113 N}{24}
          + \frac{15 ~N^2}{4}
          - N^3 $ \\[1mm]
\hline
\rule{0cm}{4mm}4 & $\frac{24367}{1152}$ & $ \frac{839}{144}
          - \frac{1601~N}{144}
          + \frac{695~N^2}{48}
          - \frac{135~N^3}{16}
          + \frac{231~N^4}{128} $ \\[1mm]
\hline
\rule{0cm}{4mm}5 & $-\frac{8821}{144}$   & $\frac{33661}{2400}
          - \frac{1151407~N}{43200}
          + \frac{197587~N^2}{4320}
          - \frac{12709~N^3}{300}
          + \frac{6271~N^4}{320}
          - \frac{7~N^5}{2} $ \\[1mm]
\hline
\end{tabular}
\caption{The coefficients $a_i$ of the leading logarithms for the mass in the
massive $O(N)$ model to five loops. Table adapted from \cite{Bijnens:2009zi}.}
\label{tab:ai}
\end{table}
The effects of the anomaly were added in \cite{Bijnens:2012hf}.
An example is the pion coupling to two off-shell photons:
\begin{eqnarray}
 A(\pi^0 \to \gamma(k_1)\gamma(k_2)) &=& \epsilon_{\mu\nu\alpha\beta}\,
\varepsilon_1^{*\mu}(k_1)\varepsilon_2^{*\nu}(k_2)\,k_1^\alpha k_2^\beta \, 
F_{\pi\gamma\gamma}(k_1^2,k_2^2)\,,
\nonumber\\ 
F_{\pi\gamma\gamma}(k_1^2,k_2^2) &=& \frac{e^2}{4\pi^2 F_\pi}\hat F
 F_\gamma(k_1^2) F_\gamma(k_2^2)
 F_{\gamma\gamma}(k_1^2,k_2^2)\,.
\end{eqnarray}
$\hat F$: is for on-shell photons; $ F_\gamma(k^2)$
is the form factor for one-off shell photon
; $ F_{\gamma\gamma}$ is the
nonfactorizable part when both photons are off-shell.
This was done to six loops. The on-shell decays leading logarithm part
converges extremely well:
\begin{equation}
 F = 1+0-0.000372+0.000088+0.000036+0.000009+0.0000002+\ldots
\end{equation}
The nonfactorizable starts only at three loops and in the massless case only at
four loops. The leading logarithms give for this a very small contribution.

The extension to the $SU(N)\times SU(N)$ case was done
in \cite{Bijnens:2013yca}. In particular we pushed $\gamma\gamma\to\pi\pi$
there to high order and found only small corrections. A summary of existing
massive leading logarithms from our work is:
\begin{itemize}
\item  $O(N)/O(N-1)$ model~\cite{Bijnens:2009zi,Bijnens:2010xg} 
\begin{itemize}
\item {massive case}: $\pi\pi$, $F_V$ and $F_S$ to 4-loop order
\item large $N$ for these cases also for massive $O(N)$.
\item done using bubble resummations or recursion equation which can be
solved analytically 
\end{itemize}
\item \cite{Bijnens:2012hf}
  \begin{itemize}
  \item $O(N)/O(N-1)$ model: Mass, $F_\pi$, $F_V$ to six loops
  \item $O(4)/O(3)$ Anomaly: $\gamma^*3\pi$ (five) and $\pi^0\gamma^*\gamma^*$ (six loops)
  \end{itemize}
\item $SU(N)\times SU(N)/SU(N)$ \cite{Bijnens:2013yca}
  \begin{itemize}
  \item Mass, Decay constants, Form-factors
  \item Meson-Meson, $\gamma\gamma\to\pi\pi$
  \end{itemize}
\end{itemize}
Typically, the expected radius of convergence was found. Large $N$ was not a
good numerical guide either to the actual coefficients unless one went to rather
large values of $N$.

Unfortunately,
in no case could we identify a conjecture for all order behaviour of leading
logarithms. I strongly recommend all of you to have a look
at the many tables in the mentioned papers to see if you have more luck there.

A final comment is that the method has recently been extended to the
 nucleon sector \cite{Bijnens:2014ila}. This is discussed in more detail in the
parallel session talk by A.A.~Vladimirov. 

\section{Conclusions}

In this talk I gave a very short introduction to Chiral Perturbation
Theory in the mesonic sector and discussed a number of recent advances.
These include the latest determination of the LECs of \cite{Bijnens:2014lea}.
Finite volume effects with twisted boundary conditions and preliminary results
on the finite volume two-loop results in three flavour ChPT were the next topic.
The third subject was the
use of mesonic ChPT and its extension to different symmetry breaking
patterns with an eye towards applications relevant to technicolour.
The last topic was the calculation of leading logarithms in a number of
effective field theories to high orders.


\begin{theacknowledgments}
I like to thank the organizers for a very pleasant conference and the
invitation to present this talk.
I also thank my collaborators whose work I have presented here and
especially J.~Vermaseren, this work would not have been
possible without {\sc FORM} \cite{FORM}.
This work is supported in part by the European Community-Research
Infrastructure Integrating Activity ``Study of Strongly Interacting Matter''
(HadronPhysics3, Grant Agreement No. 283286)
and the Swedish Research Council grants 621-2011-5080 and 621-2013-4287.
\end{theacknowledgments}



\bibliographystyle{aipproc}   


\begin{thebibliography}{9}
\bibitem{Weinberg:1978kz} 
  S.~Weinberg,
  Physica A {\bf 96}, 327 (1979).

\bibitem{Gasser:1983yg} 
  J.~Gasser and H.~Leutwyler,
  Annals Phys.\  {\bf 158}, 142 (1984).

\bibitem{Gasser:1984gg} 
  J.~Gasser and H.~Leutwyler,
  Nucl.\ Phys.\ B {\bf 250}, 465 (1985).

\bibitem{Leutwyler:1993iq} 
  H.~Leutwyler,
  Annals Phys.\  {\bf 235}, 165 (1994)
  [hep-ph/9311274].

\bibitem{chptwebpage}  \url{http://www.thep.lu.se/~bijnens/chpt/}.

\bibitem{Bijnens:1999sh} 
  J.~Bijnens, G.~Colangelo and G.~Ecker,
  JHEP {\bf 9902}, 020 (1999)
  [hep-ph/9902437].

\bibitem{Bijnens:2006zp} 
  J.~Bijnens,
  Prog.\ Part.\ Nucl.\ Phys.\  {\bf 58}, 521 (2007)
  [hep-ph/0604043].

\bibitem{Bijnens:2014lea} 
  J.~Bijnens and G.~Ecker,
  arXiv:1405.6488 [hep-ph], to be published in Ann. Rev. Nucl. Part. Sc.

\bibitem{Colangelo:2000jc} 
  G.~Colangelo, J.~Gasser and H.~Leutwyler,
  Phys.\ Lett.\ B {\bf 488}, 261 (2000)
  [hep-ph/0007112].

\bibitem{Nebreda:2012ve} 
  J.~Nebreda, J.~R.~Pelaez and G.~Rios,
  Phys.\ Rev.\ D {\bf 88}, 054001 (2013)
  [arXiv:1205.4129 [hep-ph]].

\bibitem{Aoki:2013ldr} 
  S.~Aoki {\it et al.}, 
  Eur.\ Phys.\ J.\ C {\bf 74}, no. 9, 2890 (2014)
  [arXiv:1310.8555 [hep-lat]].

\bibitem{Bijnens:1998fm} 
  J.~Bijnens, G.~Colangelo and P.~Talavera,
  JHEP {\bf 9805}, 014 (1998)
  [hep-ph/9805389].

\bibitem{Bijnens:1996wm} 
  J.~Bijnens and P.~Talavera,
  Nucl.\ Phys.\ B {\bf 489}, 387 (1997)
  [hep-ph/9610269].

\bibitem{GonzalezAlonso:2008rf} 
  M.~Gonzalez-Alonso, A.~Pich and J.~Prades,
  Phys.\ Rev.\ D {\bf 78}, 116012 (2008)
  [arXiv:0810.0760 [hep-ph]].

\bibitem{Amoros:2000mc} 
  G.~Amor\'os, J.~Bijnens and P.~Talavera,
  Nucl.\ Phys.\ B {\bf 585}, 293 (2000)
  [Erratum-ibid.\ B {\bf 598}, 665 (2001)]
  [hep-ph/0003258].

\bibitem{Amoros:2001cp} 
  G.~Amor\'os, J.~Bijnens and P.~Talavera,
  Nucl.\ Phys.\ B {\bf 602}, 87 (2001)
  [hep-ph/0101127].

\bibitem{Bijnens:2011tb} 
  J.~Bijnens and I.~Jemos,
  Nucl.\ Phys.\ B {\bf 854}, 631 (2012)
  [arXiv:1103.5945 [hep-ph]].

\bibitem{Bijnens:2004ai} 
  J.~Bijnens and F.~Borg,
  Eur.\ Phys.\ J.\ C {\bf 40}, 383 (2005)
  [hep-ph/0501163].

\bibitem{Cirigliano:2011ny} 
  V.~Cirigliano, G.~Ecker, H.~Neufeld, A.~Pich and J.~Portoles,
  Rev.\ Mod.\ Phys.\  {\bf 84}, 399 (2012)
  [arXiv:1107.6001 [hep-ph]].

\bibitem{Bijnens:2009zd} 
  J.~Bijnens and I.~Jemos,
  Eur.\ Phys.\ J.\ C {\bf 64}, 273 (2009)
  [arXiv:0906.3118 [hep-ph]].

\bibitem{Luscher:1985dn} 
  M.~Luscher,
  Commun.\ Math.\ Phys.\  {\bf 104}, 177 (1986).

\bibitem{Gasser:1986vb} 
  J.~Gasser and H.~Leutwyler,
  Phys.\ Lett.\ B {\bf 184}, 83 (1987).

\bibitem{Gasser:1987zq} 
  J.~Gasser and H.~Leutwyler,
  Nucl.\ Phys.\ B {\bf 307}, 763 (1988).

\bibitem{Golterman:2009kw} 
  M.~Golterman,
  ``Applications of chiral perturbation theory to lattice QCD,''
  arXiv:0912.4042 [hep-lat].

\bibitem{Becirevic:2003wk} 
  D.~Becirevic and G.~Villadoro,
  Phys.\ Rev.\ D {\bf 69}, 054010 (2004)
  [hep-lat/0311028].

\bibitem{DescotesGenon:2004iu} 
  S.~Descotes-Genon,
  Eur.\ Phys.\ J.\ C {\bf 40}, 81 (2005)
  [hep-ph/0410233].
\bibitem{Colangelo:2005gd} 
  G.~Colangelo, S.~Durr and C.~Haefeli,
  Nucl.\ Phys.\ B {\bf 721}, 136 (2005)
  [hep-lat/0503014].

\bibitem{Colangelo:2006mp} 
  G.~Colangelo and C.~Haefeli,
  Nucl.\ Phys.\ B {\bf 744}, 14 (2006)
  [hep-lat/0602017].

\bibitem{Bijnens:2006ve} 
  J.~Bijnens and K.~Ghorbani,
  Phys.\ Lett.\ B {\bf 636}, 51 (2006)
  [hep-lat/0602019].

\bibitem{Colangelo:2010cu} 
  G.~Colangelo, U.~Wenger and J.~M.~S.~Wu,
  Phys.\ Rev.\ D {\bf 82}, 034502 (2010)
  [arXiv:1003.0847 [hep-lat]].

\bibitem{Sachrajda:2004mi} 
  C.~T.~Sachrajda and G.~Villadoro,
  Phys.\ Lett.\ B {\bf 609}, 73 (2005)
  [hep-lat/0411033].

\bibitem{Bedaque:2004kc} 
  P.~F.~Bedaque,
  Phys.\ Lett.\ B {\bf 593}, 82 (2004)
  [nucl-th/0402051].

\bibitem{Bijnens:2014yya} 
  J.~Bijnens and J.~Relefors,
  JHEP {\bf 1405}, 015 (2014)
  [arXiv:1402.1385 [hep-lat], arXiv:1402.1385].

\bibitem{Jiang:2006gna} 
  F.-J.~Jiang and B.~C.~Tiburzi,
  Phys.\ Lett.\ B {\bf 645}, 314 (2007)
  [hep-lat/0610103].

\bibitem{BR}
  J.~Bijnens and T.~R\"ossler, to be published.

\bibitem{Bijnens:2013doa} 
  J.~Bijnens, E.~Bostr\"om and T.~A.~L\"ahde,
  JHEP {\bf 1401}, 019 (2014)
  [arXiv:1311.3531 [hep-lat]].

\bibitem{Peskin}
  M.~E.~Peskin,
  Nucl.\ Phys.\  B {\bf 175}, 197 (1980).

\bibitem{Preskill}
  J.~Preskill,
  Nucl.\ Phys.\  B {\bf 177}, 21 (1981).

\bibitem{Dimopoulos}
  S.~Dimopoulos,
  Nucl.\ Phys.\  B {\bf 168}, 69 (1980).

\bibitem{Kogut}
  J.~B.~Kogut, M.~A.~Stephanov, D.~Toublan, J.~J.~M.~Verbaarschot and A.~Zhitnitsky,
  Nucl.\ Phys.\  B {\bf 582}, 477 (2000)
  [arXiv:hep-ph/0001171].


\bibitem{Bijnens:1997vq} 
  J.~Bijnens, G.~Colangelo, G.~Ecker, J.~Gasser and M.~E.~Sainio,
  Nucl.\ Phys.\ B {\bf 508}, 263 (1997)
  [Erratum-ibid.\ B {\bf 517}, 639 (1998)]
  [hep-ph/9707291].

\bibitem{Bijnens:1995yn} 
  J.~Bijnens, G.~Colangelo, G.~Ecker, J.~Gasser and M.~E.~Sainio,
  Phys.\ Lett.\ B {\bf 374}, 210 (1996)
  [hep-ph/9511397].


\bibitem{Bijnens:2009qm} 
  J.~Bijnens and J.~Lu,
  JHEP {\bf 0911}, 116 (2009)
  [arXiv:0910.5424 [hep-ph]].

\bibitem{Bijnens:2011fm} 
  J.~Bijnens and J.~Lu,
  JHEP {\bf 1103}, 028 (2011)
  [arXiv:1102.0172 [hep-ph]].

\bibitem{Bijnens:2011xt} 
  J.~Bijnens and J.~Lu,
  JHEP {\bf 1201}, 081 (2012)
  [arXiv:1111.1886 [hep-ph]].

\bibitem{Bijnens:1999hw} 
  J.~Bijnens, G.~Colangelo and G.~Ecker,
  Annals Phys.\  {\bf 280}, 100 (2000)
  [hep-ph/9907333].

\bibitem{Bijnens:1998yu} 
  J.~Bijnens, G.~Colangelo and G.~Ecker,
  Phys.\ Lett.\ B {\bf 441}, 437 (1998)
  [hep-ph/9808421].

\bibitem{Buchler:2003vw} 
  M.~Buchler and G.~Colangelo,
  Eur.\ Phys.\ J.\ C {\bf 32}, 427 (2003)
  [hep-ph/0309049].

\bibitem{Bijnens:2009zi} 
  J.~Bijnens and L.~Carloni,
  Nucl.\ Phys.\ B {\bf 827}, 237 (2010)
  [arXiv:0909.5086 [hep-ph]].

\bibitem{Bissegger:2006ix} 
  M.~Bissegger and A.~Fuhrer,
  Phys.\ Lett.\ B {\bf 646}, 72 (2007)
  [hep-ph/0612096].


\bibitem{Kivel:2008mf} 
  N.~Kivel, M.~V.~Polyakov and A.~Vladimirov,
  Phys.\ Rev.\ Lett.\  {\bf 101}, 262001 (2008)
  [arXiv:0809.3236 [hep-ph]].

\bibitem{Kivel:2009az} 
  N.~A.~Kivel, M.~V.~Polyakov and A.~A.~Vladimirov,
  JETP Lett.\  {\bf 89}, 529 (2009)
  [arXiv:0904.3008 [hep-ph]].

\bibitem{Koschinski:2010mr} 
  J.~Koschinski, M.~V.~Polyakov and A.~A.~Vladimirov,
  Phys.\ Rev.\ D {\bf 82}, 014014 (2010)
  [arXiv:1004.2197 [hep-ph]].

\bibitem{Polyakov:2010pt} 
  M.~V.~Polyakov and A.~A.~Vladimirov,
  Theor.\ Math.\ Phys.\  {\bf 169}, 1499 (2011)
  [arXiv:1012.4205 [hep-th]].

\bibitem{Coleman:1974jh} 
  S.~R.~Coleman, R.~Jackiw and H.~D.~Politzer,
  Phys.\ Rev.\ D {\bf 10}, 2491 (1974).

\bibitem{FORM}  J.~A.~M.~Vermaseren,
  ``New features of FORM,''
  math-ph/0010025.

\bibitem{Bijnens:2010xg} 
  J.~Bijnens and L.~Carloni,
  Nucl.\ Phys.\ B {\bf 843}, 55 (2011)
  [arXiv:1008.3499 [hep-ph]].

\bibitem{Bijnens:2012hf} 
  J.~Bijnens, K.~Kampf and S.~Lanz,
  Nucl.\ Phys.\ B {\bf 860}, 245 (2012)
  [arXiv:1201.2608 [hep-ph]].

\bibitem{Bijnens:2013yca} 
  J.~Bijnens, K.~Kampf and S.~Lanz,
  Nucl.\ Phys.\ B {\bf 873}, 137 (2013)
  [arXiv:1303.3125 [hep-ph]].

\bibitem{Bijnens:2014ila} 
  J.~Bijnens and A.~A.~Vladimirov,
  arXiv:1409.6127 [hep-ph].
\end{thebibliography}



\end{document}